\documentclass[showpacs,12pt]{revtex4}

\usepackage{graphicx}%
\usepackage{dcolumn}
\usepackage{amsmath}

\begin{document}


\title{\bf Power-law behavior observed in $p_{T}$-distributions and its implications in relativistic heavy-ion collisions}


\author{MENG Ta-chung \footnote{Email address: meng@mail.ccnu.edu.cn; meng@physik.fu-berlin.de}}
\affiliation{Department of Physics, CCNU, 430079 Wuhan, China}
\affiliation{Institut f\"{u}r Theoretische Physik, FU-Berlin,
14195 Berlin, Germany}
\author{LIU Qin \footnote{Email address: liuq@mail.ccnu.edu.cn}}
\affiliation{Department of Physics, CCNU, 430079 Wuhan, China}

\date{\today}

\begin{abstract}
Preconception-free analyses of the inclusive invariant
transverse-momentum distribution data taken from the measurements
of Au+Au collisions at $\sqrt{s_{NN}}=130$ GeV and
$\sqrt{s_{NN}}=200$ GeV have been performed. It is observed that
the distributions exhibit for $p_{T}\geq 2$ GeV/c remarkably good
power-law behavior ($p_{T}$-scaling) with general regularities.
This power-law behavior leads us in particular to recognize that
the concept of centrality, albeit its simple appearance, is rather
complex; its underlying geometrical structure has to be understood
in terms of fractal dimensions. Experimental evidences and
theoretical arguments are given which show that the observed
striking features are mainly due to geometry and self-organized
criticality. A simple model is proposed which approximately
reproduces the above-mentioned data for the ``suppression''
without any adjustable parameter. Further heavy-ion collision
experiments are suggested.
\end{abstract}

\pacs{25.75.-q, 25.75.Dw, 05.65.+b, 13.85.Hd, 13.87.Ce}

\maketitle

Inclusive invariant $p_{T}$-distributions for charged hadrons in
Au+Au collisions at $\sqrt{s_{NN}}=130$ and 200 GeV have been
measured and published by STAR \cite{1,2} and by PHENIX \cite{3}
over a broad range of centrality. Such $p_{T}$-distributions for
neutral pions are also given by PHENIX \cite{4}. All these
experiments \cite{1,2,3,4} show that the hadron yields differ
appreciably at high and medium $p_{T}$ in central collisions
\begin{figure}
\includegraphics[width=0.48\textwidth]{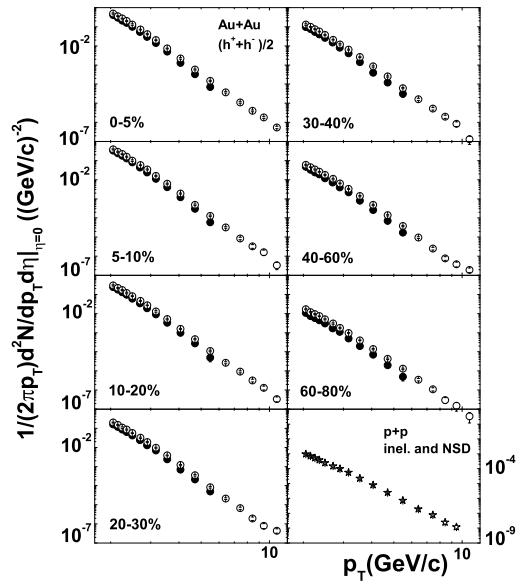}
\caption{Inclusive
invariant $p_{T}$-distribution data \cite{1,2} for $p_{T}\geq 2$
GeV/c in $\log$-$\log$ plots. Black and white indicate 130 and 200
GeV respectively.} \label{fig1}
\end{figure}
relative to peripheral collisions and to the nucleon-nucleon
reference. What do these observations, usually known as
``suppression'' \cite{5}, tell us? Are they related to the
yet-to-be-found ``quark-gluon-plasma (QGP)''? If yes, how? In
order to obtain an unbiased physical picture to start with, we
begin with preconception-free data-analyses. We then summarize the
results, and discuss their implications.

In {\it the first part} of this paper, we report on the result of
such analyses. Within the measured kinematical region $0<
p_{T}<12$ GeV/c, $\sqrt{s_{NN}}=130$ GeV and $\sqrt{s_{NN}}=200$
GeV, the inclusive invariant $p_{T}$-distributions of
$(h^{+}+h^{-})/2$ for centrality-selected Au+Au, and those for p+p
interactions exhibit power-law behavior for $p_{T}\geq 2$ GeV/c
(cf.  Fig. 1). The results can be summarized as follows.
\begin{equation}
\frac{1}{2\pi
p_{T}}\frac{d^{2}N}{dp_{T}d\eta}|_{\eta=0}(p_{T}|Au+Au;
p_{c})\propto p_{T}^{-\lambda_{AuAu}(p_{c})},
\end{equation}
\begin{equation}
\frac{1}{2\pi
p_{T}}\frac{d^{2}N}{dp_{T}d\eta}|_{\eta=0}(p_{T}|p+p)\propto
p_{T}^{-\lambda_{pp}},
\end{equation}
where the power-indices (the $\lambda$'s) are positive real
numbers, and $p_{c}$ characterizes the centrality-bins (in
percentile) which stand for the different degrees of departure
from the most central collision. The experimental values of the
$\lambda$'s obtained from the STAR data \cite{1,2} at
$\sqrt{s_{NN}}=130$ GeV and 200 GeV for different $p_{c}$-bins are
shown in Fig. 2. The results from the PHENIX data \cite{3,4} (will
be reported in a more extensive paper elsewhere \cite{6}) show
similar characteristic properties.

As we can see in Fig. 2, for a given $p_{c}$-bin, the
$\lambda$-value at $\sqrt{s_{NN}}=200$ GeV coincides with the
corresponding one at $\sqrt{s_{NN}}=130$ GeV. Furthermore, the
$\lambda$-value for the most peripheral ($p_{c}\rightarrow 100\%$)
case is very much the same as $\lambda_{pp}$'s. Note that the
$\lambda$-values increase from the most peripheral value,
$\lambda_{AuAu}(p_{c}\rightarrow100\%)\approx \lambda_{pp}$, with
decreasing $p_{c}$ to the $\lambda$-value for the most central
(center-on-certer) collision $(p_{c}\rightarrow 0\%)$ in a
monotonous manner. In this connection it is useful to consider the
ratio of both sides of Eqs. (1) and (2):
\begin{equation}
\frac{d^{2}N/p_{T}dp_{T}d\eta|_{\eta=0}(p_{T}|Au+Au;p_{c})}{d^{2}N/p_{T}dp_{T}d\eta|_{\eta=0}(p_{T}|p+p;inel.
or NSD)}\propto p_{T}^{-\lambda(p_{c})}.
\end{equation}
The quotient, which is a completely experimental quantity, on the
left-hand-side of this equation, will hereafter be referred to as
$Q(p_{T},p_{c})$; and the values of the exponent on the
right-hand-side,
\begin{equation}
\lambda(p_{c}) =\lambda_{AuAu}(p_{c})-\lambda_{pp},
\end{equation}
are dipicted in Fig. 3. They are directly obtained from the data
points shown in Fig. 2.

In {\it the second part} of this paper, we propose a simple model.
We show how the power-law behavior can be understood, and how
$\lambda_{AuAu}(p_{c})$ and $\lambda_{pp}$ in Eqs. (1) and (2) can
be estimated. The model is based on geometry and self-organized
criticality (SOC) \cite{7,8}. As we shall see, both geometry and
SOC contribute powers in $p_{T}$ to the distributions shown in
Eqs. (1) and (2). The relevant facts and arguments are given
below:
\begin{figure}
\includegraphics[width=0.46\textwidth]{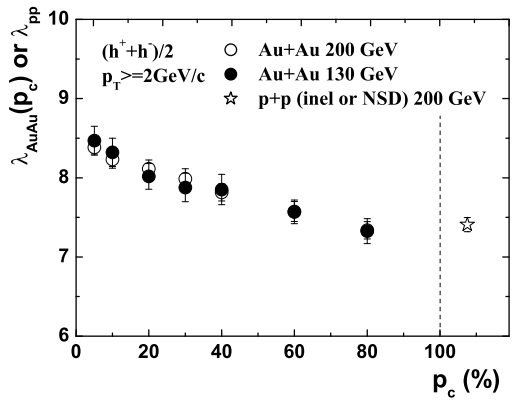}
\caption{The power-indices, $\lambda_{AuAu}(p_{c})$ and
$\lambda_{pp}$, evaluated by measuring the slopes in Fig. 1 are
plotted as function of $p_{c}$.} \label{fig2}
\end{figure}

(A) Geometry: Let us first recall the well-known observation made
by Rutherford \cite{9} on large-momentum-transfer scattering, and
a less-known observation made by Williams \cite{10} in which the
following has been pointed out: Ordinary space-time concepts are
useful for the semiclassical description of high-energy scattering
processes, {\it provided that} the de Broglie wavelength of the
projectile is short compared to the linear dimension of the
scattering field, {\it and provided that} the corresponding
momentum transfer which determines the deflection is not smaller
than the disturbance allowed by the uncertainty principle. Through
a simple realistic estimation, we can, and we did, convince
ourselves that all these conditions {\it are indeed fulfilled} for
Au+Au and p+p collisions at $\sqrt{s_{NN}}\geq 130$ GeV and
$p_{T}\geq 2$ GeV/c. Furthermore we note that the corresponding
phase-space factors can be estimated by making use of the
uncertainty principle in accordance with Refs. \cite{9} and
\cite{10}.

(B) SOC: It is well-known that approximately $50\%$ of the kinetic
energy of every fast moving hadron is carried by its gluonic
content and that the characteristic properties of the gluons, in
particular, the direct gluon-gluon coupling prescribed by the QCD
Lagrangian, the confinement, and the nonconservation of gluon
numbers, can and should be considered as more than a hint that
systems of interacting soft gluons are open dynamical complex
systems which are far from thermal and/or chemical equilibrium.
Taken together with the observations \cite{7,8} made by Bak, Tang,
and Wiesenfeld (BTW), these facts strongly suggest the existence
of SOC and thus the existence of BTW avalanches in gluonic systems
\cite{11,12}.

According to SOC, a small part of such BTW avalanches manifests
themselves in the form of color-singlet gluon clusters
$c_{0}^{\star}$, and that they can be readily examined
\cite{11,12} experimentally in inelastic diffractive scattering
processes \cite{13}. This is because the interactions between the
\begin{figure}
\includegraphics[width=0.47\textwidth]{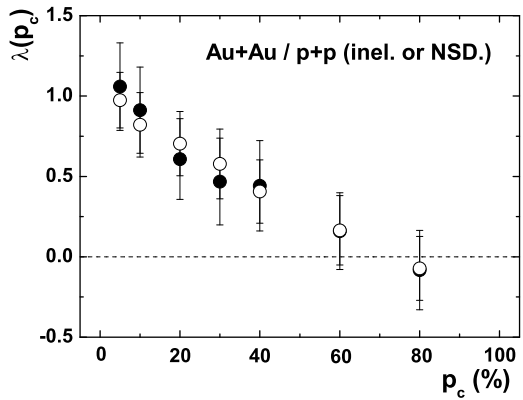}
\caption{The power-index $\lambda (p_{c})$ defined in Eq. (4)
plotted as function of $p_{c}$. Data are from Refs. \cite{1,2}.}
\label{fig3}
\end{figure}
struck $c^{\star}_{0}$ and any other color singlets are of Van der
Waal's type which are much weaker than color forces at distances
of the order of hadron radius. In fact, in order to check the
existence and the properties of the $c^{\star}_{0}$'s, a
systematic data analysis has been performed \cite{12}, the result
of which shows that the size distribution $D_{S}(S)$, and the
lifetime distribution $D_{T}(T)$ of such $c^{\star}_{0}$'s indeed
exhibit power-law behavior $D_{S}(S)\propto S^{-\mu}$, $D_{T}(T)
\propto T^{-\nu}$, where $\mu$ and $\nu$ are positive real
constants. Such characteristic features are known as ``the
fingerprints of SOC'' \cite{7,8}. These fingerprints imply in
inelastic diffractive scattering, the size S of the struck
$c^{\star}_{0}$ is proportional to the directly measurable
quantity $x_{P}$, which is the energy fraction carried by ``the
exchanged colorless object'' in such processes, the existing data
\cite{13} show $D_{S}(x_{P}) \propto x_{p}^{-\mu}$, where $\mu
=1.95\pm 0.12$ \cite{11,12}.

By considering inelastic diffractive scattering \cite{11,12}, we
were able to check---and only able to check---the existence and
properties of the color-singlet gluon clusters. But, due to the
observed SU(3) color symmetry, most of such gluon clusters are
expected to be color multiplets which will hereafter be denoted by
$c^{\star}$'s. Furthermore, in accordance with the experimentally
confirmed characteristic features of the BTW theory, the SOC
fingerprints in gluon systems should not depend on the dynamical
details of their interactions, in particular, not on the details
about the exchanged quantum numbers in their formation processes.
This implies that $D_{S}(S)$ and $D_{T}(T)$ of the $c^{\star}$'s
are expected to have \cite{14} not only the same power-law
behavior but also the same power as that of their color-singlet
counterparts observed in inelastic diffractive scattering
processes \cite{11,12}.

The fact \cite{13} that quarks can be knocked out of protons by
projectiles whenever large-momentum-transfer between projectiles
and targets take place, has prompt us to propose \cite{14} that
$c^{\star}$'s can also be ``knocked out'' of the mother proton by
a projectile provided that the corresponding transfer of momenta
is large enough where the knocked-out $c^{\star}$'s may have
``color-lines'' connected to the remnant of the proton. What we
show now is that the observed power-law behavior in Eq. (2) can be
quantitatively described by the product of the probability
distribution(s) of the knocked-out $c^{\star}$'s and the
phase-space factors associated with the knock-out processes.

Recall that processes of inclusive high-$p_{T}$ jet-production,
p+p$\rightarrow$jet+$X$, at high energies are dominated by two-jet
events; and that in a SOC-based model \cite{14}, such jets are
produced in two-step-processes. In Step 1: A quark $q$ ($q_{v}$ or
$q_{s}$ or $\bar{q_{s}}$) in one of the colliding nucleon collides
with a quark $q$ ($q_{v}$ or $q_{s}$ or $\bar{q_{s}}$) in the
other nucleon where an amount of $p_{T}$ is transferred in the
plane in which the two nucleons in form of thin contracted objects
meet each other, and in which large-$p_{T}$ quark-quark scattering
takes place. In Step 2: Since the two scattered $q$'s and/or
$\bar{q}$'s are in general space-like (because of the large
$p_{T}$), the easiest way for them to escape the confinement
region is each ``catches'' a suitable time-like (in order to
provide the high-$p_{T}$ $q$ or $\bar{q}$ with sufficient energy)
anticolor BTW-avalanches, $c^{\star}$'s, which in accordance with
the SOC-picture exist in abundance in their neighborhood. This is
how a color-singlet jet is created. A scattered $q$ or $\bar{q}$
can also combine with a colored $c^{\star}$ to form a jet or a fan
which is connected with other colored object(s) through
color-lines. This is how color multiplet jets (or fans) can be
produced. Hence, in the proposed picture the invariant cross
section $Ed^{3}\sigma/dp^{3}$ is expected to have the following
factors.

(i) A phase-space factor that describes the chance for the two
quarks ($q$ or $\bar{q}$, $\cdots$) which initiate Step 1 to come
so close to each other in space that they can exchange a large
$p_{T}$ ($\approx E_{T}$). This phase-space factor can be
estimated by making use of the uncertainty principle and the two
observations mentioned in (A) above. By choosing the $z$ axis as
the collision axis, $p_{T}$ will be in (or near) the $xy$ plane.
The chance for two constituents ($q$ or $\bar{q}$, $\cdots$)
moving parallel to the $z$ axis to come so close to each other in
the $xy$ plane such that an amount $p_{T}$ can be transferred is
approximately proportional to the size of the corresponding phase
space $\Delta x\Delta y \sim (\Delta p_{x})^{-1}(\Delta
p_{y})^{-1}\sim p_{T}^{-1}p_{T}^{-1}= p_{T}^{-2}$.

(ii) Since each jet is associated with a to be knocked-out gluon
cluster $c^{\star}$, which has energy comparable with $E_{T}$,
$Ed^{3}\sigma/dp^{3}$ is expected to be proportional to the square
of the probability $D_{S}$ to find such a $c^{\star}$. The size S
of BTW avalanches is directly proportional to $x_{p}$, thus
proportional to the energy $E_{T}$ it carries. Furthermore, since
$E_{T}\approx p_{T}$ for high-energy jets \cite{14}, we have:
\begin{equation}
D_{S}(x_{p})\propto D_{S}(E_{T})\propto p_{T}^{-\mu}.
\end{equation}
This means, we expect to see a factor $p_{T}^{-2\mu}$ in the
invariant cross section $Ed^{3}\sigma/dp^{3}$.

(iii) Having in mind that the scattered quarks are in general
space-like, two-step-processes are expected to take place only
when there are suitable $c^{\star}$'s in the surroundings
immediately after the first step. The probability of having
sufficient $c^{\star}$'s around, wherever and whenever two
constituents meet during the p+p collision, is guaranteed when the
c.m.s of one proton meets that of the other. Hence, phase-space
considerations w.r.t time requires a factor $(\Delta t)^{2}\sim
E_{T}^{-2}\sim p_{T}^{-2}$.

Hence, for p+p collisions, we expect to see $Ed^{3}N/dp^{3}\propto
Ed^{3}\sigma/d^{3}p\propto p_{T}^{-2-2-2\mu}$. By taking the lower
limit of $\mu $, we obtain:
\begin{equation}
\frac{1}{2\pi
p_{T}}\frac{d^{2}N}{dp_{T}d\eta}|_{\eta=0}(p_{T}|p+p)\propto
p_{T}^{-7.66}
\end{equation}
which is in reasonable agreement with the data \cite{2}.

Next, we focus our attention to the {\it empirical} result
described by Eqs. (3) and (4) together with Fig. 3. Note that
according to Eq. (3), the quotient $Q(p_{T},p_{c})$ stands for the
chance to find a large-$p_{T}$ charged hadron in Au+Au collisions
within a given $p_{c}$ range; and this chance is measured in
``units'' of the chance in finding a similar large-$p_{T}$ hadron
in p+p collisions. We now take a closer look at the straight
lines, on which the data-points lie in the log-log plots of Fig.
1, and note the fact that the slopes depend only on
$p_{c}$---independent of $\sqrt{s_{NN}}$. This has to be
considered as a strong hint at the possible distinguished role
played by geometry in describing/understanding such collision
processes. Hence, it is useful not only to recall the facts and
the arguments mentioned in (A) and those in (i) above, but also to
recall that the word ``centrality'' $p_{c}$ is in fact very much
involved: Experimentally \cite{1,3} it is determined by measuring
the multiplicities of produced hadrons; but, since the notion of
``{\it departure from the center-on-center case}'' is geometrical,
it is expected to be describable in terms of geometry, in
particular in terms of impact parameters, $b$'s. These facts and
arguments have led us to propose the following picture.
High-$p_{T}$ jet-production processes in relativistic heavy-ion
(AA) collisions can be viewed as an ensemble of corresponding
jet-production processes in binary nucleon-nucleon (NN)
collisions. The observed effects depend significantly on
collision-geometry.

Since every AA event corresponds to a $b$-parameter (note that the
reverse is not true), an ensemble of collision events corresponds
to a set of $b$-parameters. By choosing the $z$-axis as the
collision axis where the centers of the two colliding nuclei are
located at (-b/2,0) and (b/2,0), on the $x$-axis of the $xy$-plane
in every event, we obtain point sets of impact parameters. It is
on such point sets, {\it the geometrical support}, we study the
above-mentioned $p_{T}$-distributions. Having the observations
made by Rutherford \cite{9} and Williams \cite{10} in mind [cf.
(A) and (i) above], the relation $\Delta x\sim (\Delta
p_{x})^{-1}\sim p_{T}^{-1}$ obtained by using the uncertainty
principle tells us the following. For every measured value
$p_{T}$, there is an interval $\Delta x$; and it is with {\it this
precision} in the corresponding spatial coordinate that the
probability $Q(p_{T},p_{c})$ [precisely speaking $Q(\Delta
x,p_{c})$ on its geometrical support] of finding high-$p_{T}$
charged hadrons can be measured.

In the proposed picture based on SOC and geometry, we are not (at
least not yet) in a position to make predictions for $Q(\Delta
x,p_{c})$ or its geometrical support---not even the dimensions of
such object! But fortunately, we know how to measure them! Thank
the master-mathematicians: K. Weierstrass, G. Cantor, H.von Koch,
F. Hausdorff, $\cdots$, P. L\'{e}vy and B. Mandelbrot \cite{15},
we learned how to use the box-counting technique. Due to the facts
and the arguments given in (A) and (i) above, we know that the
length of the boxes in our case are of the order $\Delta x\sim
(\Delta p_{x})^{-1}\sim p_{T}^{-1}$ which implies: $\Delta x$
becomes smaller and smaller for larger and larger $p_{T}$. Hence
the observed $p_{T}$-scaling  tells us that the result of this
box-counting is nothing else but the result summarized in Eq. (3)
which can also be written as $Q(\Delta x,p_{c})\propto (\Delta
x)^{\lambda(p_{c})}$. Since this observation is independent of the
positions of the boxes in each $p_{c}$-bin, by normalizing the
probabilities $Q(\Delta x,p_{c})$, the number of boxes, $N(\Delta
x,p_{c})$, needed to cover the produced hadrons distributed on the
geometrical support is proportional to the inverse of $Q(\Delta
x,p_{c})$. That is: $N(\Delta x,p_{c})\propto (\Delta
x)^{-\lambda(p_{c})}$. Hence, in the limit of large $p_{T}$, thus
small $\Delta x$, $\lambda(p_{c})$ is the corresponding fractal
dimension of the geometrical support which consists of set of
impact parameter points within each given $p_{c}$-bin. This means,
in the proposed model, we expect to see that the inclusive
invariant $p_{T}$-distributions for $p_{T}\geq 2$ GeV/c in any
kind of relativistic heavy-ion (AA) collisions satisfy
\begin{equation}
\frac{1}{2\pi p_{T}}\frac{d^{2}N}{dp_{T}d\eta}|_{\eta
=0}(p_{T}|AA;p_{c})\propto p_{T}^{-\lambda_{NN}-\lambda(p_{c})}
\end{equation}
where $\lambda_{NN}\approx7.66$, and the $\lambda(p_{c})$'s are
the fractal dimensions of the point-set of impact parameters.

It would be exciting to see further experiments with larger
$p_{T}$-values, at higher energies and with other kinds of
colliding nuclei. Such experiments will not only serve the general
goal of QGP-search, but also, in particular, be able to check
whether the empirical regularities are indeed as general as the
existing data seem to suggest, and to check whether/how concepts
and methods of Complex Sciences in particular those borrowed from
Fractal Geometry and SOC are helpful in understanding Relativistic
Heavy-Ion Physics.

The authors thank XU Nu for explaining the experiments of STAR
Collaboration to us. We thank LIANG Zuo-tang and XU Nu for their
critical remarks and useful suggestions. We also thank LAN Xun and
LIU Lei for helpful discussions and KeYanChu of CCNU for financial
support. This work is supported by the National Natural Science
Foundation of China under Grant No. 70271064.

\bibliography{apssamp}

\begin{thebibliography}{s2}
\bibitem{1}
C. Adler {\it et al.}, Phys. Rev. Lett. {\bf 89}, 202301 (2002).

\bibitem{2}
J. Adams {\it et al.}, Phys. Rev. Lett. {\bf 91}, 172302 (2003).

\bibitem{3}
K. Adcox {\it et al.}, Phys. Rev. Lett. {\bf 88}, 022301 (2002).

\bibitem{4}
S.S. Adler {\it et al.}, Phys. Rev. Lett. {\bf 91}, 072301 (2003).

\bibitem{5}
The precise definition of ``suppression'' can, e.g., be found in
Refs \cite{1,2,3,4} and the references herein.

\bibitem{6}
Q. Liu, L. Liu, X. Lan and T. Meng (in preparation).

\bibitem{7}
P. Bak, C. Tang, and K. Wiesenfeld, Phys. Rev. Lett. {\bf 59}, 381
(1987); P. Bak, C. Tang, and K. Wiesenfeld, Phys. Rev. A {\bf 38},
364 (1988).

\bibitem{8}
P. Bak, {\it How Nature Works} (Springer, New York, 1996); H. J.
Jensen, {\it Self-Organized Criticality} (Cambridge University
Press, Cambridge, UK, 1998).

\bibitem{9}
E. Rutherford, Philos. Mag. {\bf XXI}, 669 (1911).

\bibitem{10}
E. J. Williams, Rev. Mod. Phys. {\bf 17}, 217 (1945).

\bibitem{11}
T. Meng, R. Rittel, and Y. Zhang, Phys. Rev. Lett. {\bf 82}, 2044
(1999).

\bibitem{12}
C. Boros, T. Meng, R. Rittel, K. Tabelow, and Y. Zhang, Phys. Rev.
D {\bf 61}, 094010 (2000).

\bibitem{13}
See, e.g., H. Abramowicz and A. Caldwell, Rev. Mod. Phys. {\bf
71}, 1275 (1999); A. M. Cooper-Sarkar, R.C. E. Devenish, and A. de
Roeck, Int. J. Mod. Phys. A {\bf 13}, 3385 (1998), and references
therein.

\bibitem{14}
J. Fu, T. Meng, R. Rittel and K. Tabelow, Phys. Rev. Lett. {\bf
86}, 1961 (2001).

\bibitem{15}
B.B. Mandelbrot, Science {\bf 155}, 636 (1967). B.B. Mandelbrot,
{\it The Fractal Geometry of Nature}, W.H. Freeman, N.Y. G.A.
Edgar, {\it Classics on fractals}, Westoiew, 2004.

\end{thebibliography}


\end{document}